\documentclass[pre,twocolumn]{revtex4}

\usepackage{graphicx}
\usepackage{dcolumn}
\usepackage{bm}
\usepackage{amsmath}
\usepackage{textcomp}

\newcommand{\be}{\begin{equation}}
\newcommand{\ee}{\end{equation}}

\begin{document}

\title{Observation of localized modes at phase slips in two-dimensional photonic lattices}

\author{Alexander Szameit$^{1}$, Mario I. Molina$^{2,3}$, Matthias Heinrich$^4$\\Felix Dreisow$^4$, Robert Keil$^4$, Stefan Nolte$^4$, and Yuri S. Kivshar$^5$}

\affiliation{
$^1$Solid State Institute and Physics Department, Technion, 32000 Haifa, Israel\\
$^2$Departmento de F\'{\i}sica, Facultad de Ciencias, Universidad de Chile, Santiago, Chile\\
$^3$Center for Optics and Photonics, Universidad de Concepci\'{o}n, Casilla 4016, Concepci\'{o}n, Chile\\
$^4$Institute of Applied Physics, Friedrich-Schiller-Universit\"{a}t Jena, Max-Wien-Platz 1, 07743 Jena, Germany\\
$^5$Nonlinear Physics Center, Research School of Physics and Engineering, Australian National University, Canberra ACT 0200,
Australia}

%\date{\today}

\begin{abstract}
We study experimentally light localization at phase-slip waveguides
and at the intersection of phase-slips in a two-dimensional (2D)
square photonic lattice. Such system allows to observe a variety of
effects, including the existence of spatially localized modes for
low powers, the generation of strongly localized states in the form
of discrete bulk and surface solitons, as well as a crossover between one-dimensional (1D) and 2D
localization.
\end{abstract}

%\ocis{190.4420; 190.5530; 190.4350}

\maketitle

The study of nonlinear dynamics in waveguide arrays and photonic
lattices has attracted a great deal of attention due to the
possibility to observe many novel effects of nonlinear physics and
possible interesting applications in photonics~\cite{review}. In
particular, it was shown that discrete nonlinear photonic systems
can support different types of spatially localized states in the
form of discrete solitons~\cite{Christodoulides:DiscreteSoliton}.
These solitons can be controlled by the insertion of suitable
defects in an array, as was suggested theoretically~\cite{krolik}
and also verified experimentally for arrays of optical
waveguides~\cite{exp_mor}. Defects may provide an additional
physical mechanism for light confinement, and they can support
spatially localized modes, which have been studied theoretically for
different linear~\cite{tromp_2003} and nonlinear~\cite{kiv,sukh,exp_chen}
models and observed experimentally in
1D~\cite{tromp_2003,chen2} and
2D~\cite{Szameit:SurfaceDefectSolitons} photonic lattices.

\par Recently, a novel type of nonlinear mode has been
introduced~\cite{our_ol,our_pra} at the interface between two
spatially shifted nonlinear waveguide arrays. These
defect modes are closely linked with the 
phase-slip defects in 2D photonic crystals~\cite{raikh}. In the
presence of a phase-slip, the distance between two lattice sites
located at both sides of the phase-slip is a non-integer multiple of
the lattice constant. As a result, the nonlinear phase-slip defect
modes possess the specific properties of both discrete nonlinear surface
modes and bulk solitons.

\par In this Letter we study experimentally, for the first time to our
knowledge, light localization in a 2D square photonic lattice
containing one phase-slip defect or an intersection of two
phase-slip defects. We observe a variety of novel effects, including
linear modes localized at the cross intersection of the
phase-slips~\cite{raikh}, strongly nonlinear localized states in the
form of discrete solitons at a single slip~\cite{our_ol}, and
discrete surface solitons at the lattice edges~\cite{our_pra}. We
also observe and discuss a crossover between 1D and 2D localization.

\par The considered system consists of an array of $N\times N$ nonlinear
focussing (Kerr) waveguides, originally forming a square lattice
with periodicity $a$. Breaking the 
%%%%%%%%%%%%%%%%%%%
\begin{figure}[t]
\includegraphics[width=0.75\linewidth]{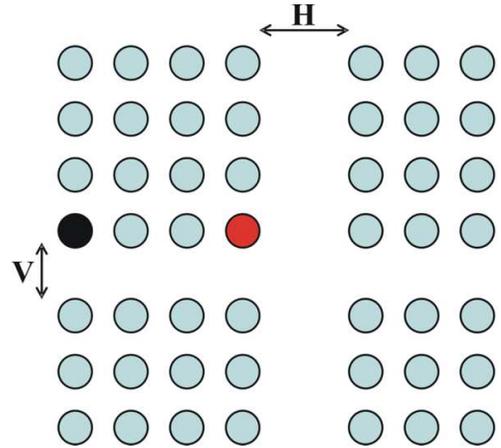}
\caption[argumento]{(Color online) Schematics of an intersection of two different phase
slips ($H,V \neq a$). The case of a single phase slip corresponds to
either $H \neq a$ or $V \neq a$. Marked waveguides show the excitation
points.}
 \label{fig1}
\end{figure}
%%%%%%%%%%%%%%%%%%%%%
translational
symmetry of the system by altering the distance between
two consecutive rows or columns creates a phase defect or a cross
intersection of two phase-slips (see Fig. \ref{fig1}). The coupling across
the defect in the x-direction (y-direction) is termed
$C_{H}(C_{V})$. In the framework of the couple-mode theory, the
evolution equations for the electric field amplitudes $E_{n}(z)$
can be written in the form,

\be
  i \frac{dE_{n}}{dz} + \sum_{m} C_{n,m} E_{m} + \gamma |E_{n}|^{2}
  E_{n}=0 \label{eq:0}
\ee

where $E_{n}(z)$ is the electric field (in units of
$(\mbox{Watt})^{1/2}$) on waveguide $n$ at distance $z$ (in
$\mbox{meters}$), $C_{n,m}$ is the coupling coefficient
(in units of $1/\mbox{meter}$) between guides $n$ and $m$, and $\gamma$ (in units of $1/(\mbox{Watt}\times \mbox{meter}$))
is the nonlinear coefficient, defined by $\gamma = \omega_{0}
n_{2}/c A_{\mathrm{eff}}$, where $\omega_{0}$ is the angular frequency
of light, $n_{2}$ is the nonlinear refractive index of the guide and
$A_{\mathrm{eff}}$ is the effective area of the linear modes. Equation
(\ref{eq:0}) is normalized by defining a
dimensionless distance $s\equiv C z$, where $C$ is the coupling between
nearest-neighbor guides outside the vicinity of the slip defect,
and the dimensionless electric field $\phi_{n}=(\gamma/C)^{1/2}
E_{n}$. With the above definitions, the conserved power (in
Watts) is given by $\sum_{n} |E_{n}|^{2}= (C/\gamma)\ P$,
%%%%%%%%%%%%%%%%
\begin{figure}[t]
\includegraphics[width=0.45\linewidth]{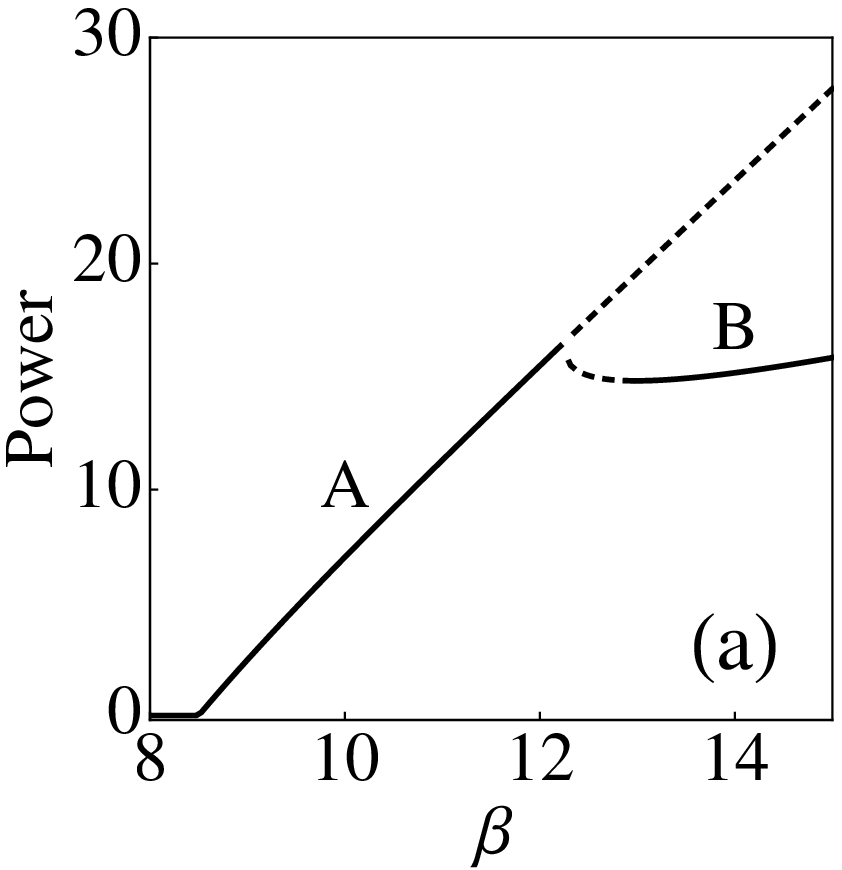}
\includegraphics[width=0.45\linewidth]{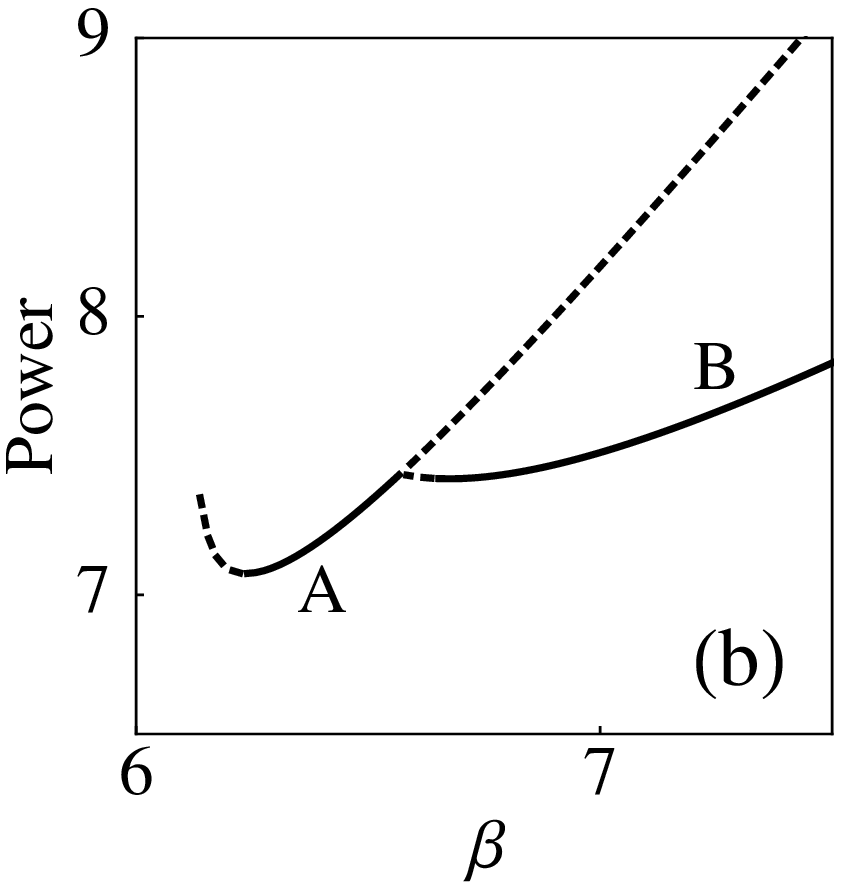}
\caption[argumento]{Power diagrams for the cases: (a) The mode is
localized at a cross intersection of two phase slips with $V=H=$29
\textmu m. (b) Surface mode localized at a single phase slip with $V=
$25 \textmu m, $H=a=$ 40 \textmu m. Solid lines refer to
stable states while dashed branches refer to the unstable modes.
} \label{fig2}
\end{figure}
%%%%%%%%%%%%%%%%
where $P\equiv \sum_{n} |\phi_{n}|^{2}$ is the dimensionless power.
As was first shown by Apalkov and Raikh~\cite{raikh}, a topological
defect
%%%%%%%%%%%%%%%%
\begin{figure}[!h]
\includegraphics[width=0.90\linewidth]{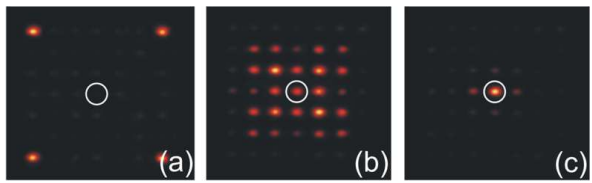}
\caption[argumento]{(Color online) Light diffraction and
localization in a perfect square lattice, for different input
powers: (a) 75 kW, (b) 1600 kW, and (c) 2000 kW. Excited waveguides
are marked with white circles.} \label{fig3}
\end{figure}
%%%%%%%%%%%%%%%%
 created by the intersection of two phase-slips, can support a
{\em linear} localized mode. Looking for stationary {\em linear}
modes ($\gamma=0$) of the system (\ref{eq:0}) in the form,
$\phi_{n}(s) = \phi_{n} \exp(i \beta s)$, we find a linear
localized mode at a fixed value of the propagation constant,

\be
  \beta = C_{V} + {1\over{C_{V}}}  + C_{H} + {1\over{C_{H}}},
\ee

provided $C_{V}>1$ and $ C_{H}>1$. In the nonlinear case, we follow the recent theoretical
results~\cite{our_ol,our_pra} and demonstrate that nonlinearity can
support a variety of localized modes including the modes which
bifurcate from the symmetric states and describe nonlinear
asymmetric localized states. The corresponding bifurcation diagram
of low-power branches of localized modes in shown in Figs.~\ref{fig2}(a,b)
for bulk and surface modes, respectively. When the mode power grows,
the symmetric mode A described in the anti-continuum limit by the
pattern $(\begin{smallmatrix}+ & + \\ +& + \end{smallmatrix})$ becomes unstable,
and it transforms into the asymmetric mode B with the pattern
$(\begin{smallmatrix}+ & 0\\0 & 0\end{smallmatrix})$ corresponding to the lower
bifurcation branch in Fig.~\ref{fig2}(a). Similarly, the surface symmetric
mode $(\begin{smallmatrix}+\\+\end{smallmatrix})$ bifurcates into the
asymmetric mode $(\begin{smallmatrix}+\\0\end{smallmatrix})$  above a certain
power threshold, as shown in Fig.~2(b). Whereas in the case of
Fig.~\ref{fig2}(a), the localized modes extend toward vanishing powers, a
finite threshold power is required to generate a surface mode in the
case of Fig.~\ref{fig2}(b).

For the fabrication, we employ the femtosecond
direct-writing technique~\cite{itoh}. The specific fabrication
parameters can be found, e.g., in~\cite{szameit06}. Each array is
100 mm long and consists of 7 $\times$ 7 waveguides, as shown in
Fig. 1(a). The spacing between the individual lattice sites is
$a=$40\textmu m except in the phase-slip channels, where the spacing is
decreased. In the experiments, light of a Ti:sapphire chirped pulse
amplification laser system (RegA, Coherent), with a pulse
duration of about 180 fs and a repetition rate of 150 kHz at 800 nm,
was coupled into the central guide using a 4 $\times$ microscope
objective [numerical aperture (NA) of 0.1], coupled out by a 10
$\times$ objective (NA = 0.25), and projected onto a CCD camera.

First, we study the light propagation in a perfect lattice $V=H= a=$40\textmu m. Figures \ref{fig3}(a-c) show the
experimental images of the light at the lattice output for different
input powers. In the linear regime [Fig. \ref{fig3}(a)], we observe strong discrete
diffraction but, when the input power grows, the diffraction is
suppressed [Fig. \ref{fig3}(b)], and the generation of a discrete 2D
soliton~\cite{fleischer} is observed for the input power of 2000 kW [Fig. \ref{fig3}(c)].

Next, we consider a symmetric cross-intersection of 
%%%%%%%%%%%%%%%%%%%%%%%
\begin{figure}[!h]
\includegraphics[width=0.90\linewidth]{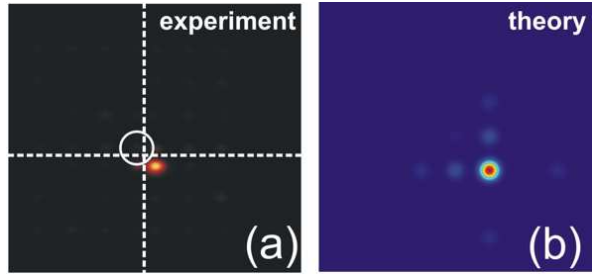}
\caption[argumento]{(Color online) (a) Experimental and (b)
Theoretical results for the generation of a linear localized mode at
an intersection of two phase-slip waveguides for $V=H=$22 \textmu m and
an input power 75 kW [cf. Fig.~3(a)]. Excited waveguides are marked
with white circles; phase slips are marked by dashed lines.}
\label{fig4}
\end{figure}
%%%%%%%%%%%%%%%%%%%%%%%
two phase-slip
waveguides with the parameters $V=H=$22 \textmu m, and excite one of the
waveguides near the cross, as marked in Fig.~\ref{fig1}. Even for a power
level as low as 75 kW we observe, in a sharp contrast with the
diffraction pattern of Fig. \ref{fig3}(b), the generation of a linear mode
localized at the phase-slip intersection. Figures \ref{fig4}(a,b) show both
experimental and numerical images of the light intensity at the
output facet that confirm the generation of a strongly localized state
at the phase-slip defect, in accord with the theory~\cite{raikh}. Because we excite only one site of the
lattice, the resulting four-site mode can not be generated as a
stationary mode, but it is clearly seen how the light is bound to the phase-slip, which is reproduced
by direct numerical simulations, see Fig. \ref{fig4}(b).

For the next series of experiments, we fabricate a single phase slip
and excite one of the central 
%%%%%%%%%%%%%%%%%%%%%%%%%%%%%%
\begin{figure}[t]
\includegraphics[width=0.90\linewidth]{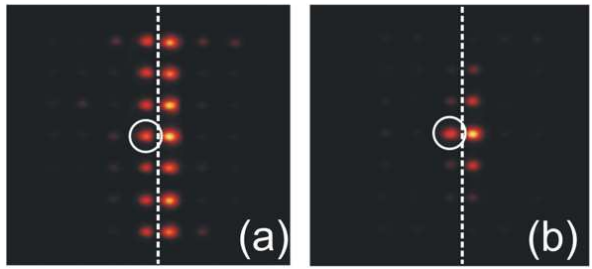}
\caption[argumento]{(Color online) Experimental images of the light
intensity at the output facet showing the generation of (a) a
1D stripe surface soliton for an input power 1700 kW,
and (b) its transformation into a 2D soliton at 2000
kW. The waveguide parameters are: for $H=$29 \textmu m, $V=a=$40 \textmu m.
Excited waveguides are marked with white circles; phase slips are marked by dashed lines.} \label{fig5}
\end{figure}
%%%%%%%%%%%%%%%%%%%%%%%%%%%%%%
%%%%%%%%%%%%%%%%%%%%%%%%%%%%%%
\begin{figure}[!h]
\noindent
\includegraphics[width=0.90\linewidth]{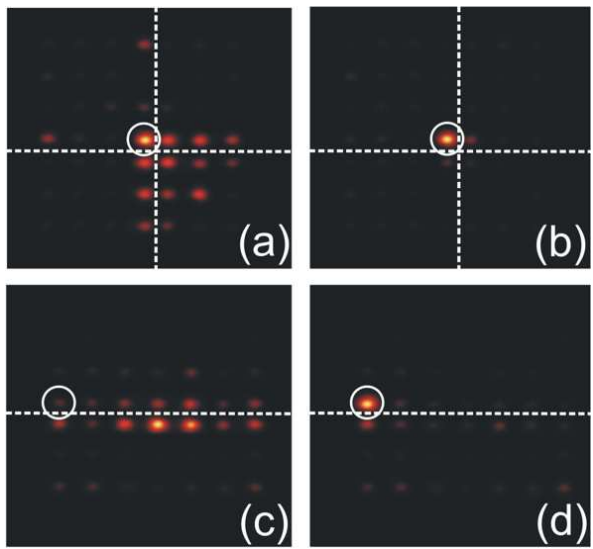}
\caption[argumento]{(Color online) Experimental images of the output
light intensity showing the generation of (a,b) 2D
soliton ($H=V=$29\textmu m), and (c,d) generation of a surface soliton at
a phase slip  ($H=$40 \textmu m and $V=$25 \textmu m). Excited waveguides are marked
with white circles; phase slips are marked by dashed lines.} \label{fig6}
\end{figure}

%%%%%%%%%%%%%%%%%%%%%%%%%%%%%%
waveguides, as
shown in Fig.~\ref{fig5}(a). For low input power, we excite one of the
linear guided modes of this phase slip waveguides (not shown), that
undergoes symmetry breaking for larger powers and transforms into an
asymmetric 1D stripe soliton, as shown in Fig.~\ref{fig5}(a) for 1700 kW. Increasing the input power further, we
observe a sharp crossover between 1D and 2D
localization and the generation of a 2D soliton, see Fig.~\ref{fig5}(b).

The generation of a 2D asymmetric nonlinear mode at the
intersection of two phase-slips ($H=V=$29 \textmu m) is shown in
Fig.~\ref{fig6}(a,b). Even at relatively large power (1400 kW), the light
diffracts across a quarter of the lattice [Fig. \ref{fig6}(a)], but for higher powers
(2500 kW) the mode becomes localized eventually [Fig. \ref{fig6}(b)].

An example for nonlinear surface mode is
presented in Fig.~\ref{fig6}(c,d) for a single phase-slip
waveguide, $H=$40 \textmu m and $V=$25 \textmu m. For lower powers (75 kW),
the light is repelled by the surface [Fig. \ref{fig6}(c)]. This behavior is
very similar to the light propagation in one-dimensional waveguide
arrays~\cite{arrays_surface}. However, when the input power exceeds
a threshold (2800 kW), we observe the generation of a localized
state at the surface [Fig. \ref{fig6}(d)].

In conclusion, we have studied experimentally the generation of
spatially localized modes at phase-slip waveguides and their
intersections. We have
generated both discrete bulk and surface solitons near the lattice
structural defects, in a qualitative agreement with earlier
theoretical predictions. We have also observed an interplay between
the effectively 1D and 2D dynamics.

This work was supported by the Leopoldina-German Academy of Science
(grant LPDS 2009-13), Fondecyt (grant 1080374), Programa de
Financiamiento Basal de Conicyt (grant FB0824/2008), Deutsche
Forschungsgemeinschaft (Leibniz program), and by the Australian
Research Council.

\end{document}